\documentstyle[aps,prd,epsfig]{revtex}

\newcommand{\ve}{\varepsilon}
\def\thrhalf{\mbox{\tiny{$3/2$}}}
\def\half{\mbox{\small{$\frac{1}{2}$}}}
\newcommand{\be}{\begin{equation}}
\newcommand{\ee}{\end{equation}}
\newcommand{\tr}{{\tilde r}}

\newcommand{\integ}{\int_{0}^{\infty}}

\begin{document}

\draft

\title{Mass and Weak Field Limit of Boson Stars in Brans Dicke Gravity}
\author{A. W. Whinnett\footnote{Email: A.W.Whinnett@qmw.ac.uk}
        \\School of Mathematical Sciences
        \\Queen Mary and Westfield College
        \\University of London}

\maketitle

\begin{abstract}

We study boson stars in Brans Dicke gravity and use them to
illustrate some of the properties of three different mass definitions: the
Schwarzschild mass, the Keplerian mass and the Tensor mass. We analyse
the weak field limit of the solutions and show that only
the Tensor mass leads to a physically reasonable definition of 
the binding energy.  We examine numerically strong field $\omega=-1$
solutions and show how, in this extreme case, the three mass values
and the conserved particle number behave as a function of the central
boson field amplitude. The numerical studies imply that for
$\omega=-1$, solutions with
extremal Tensor mass also have extremal particle number. This is a
property that a physically reasonable definition of the mass of a
boson star must have, and we prove analytically that this is true for
all values of $\omega$. The analysis supports the conjecture that the
Tensor mass uniquely describes the total energy of an asymptotically
flat solution in BD gravity.

\end{abstract}

\pacs{PACS numbers: 04.40.Dg, 04.50.+h}
\vskip2pc]


\section{Introduction}\label{intro}

Boson stars (or Klein Gordon geons) are asymptotically flat self gravitating
configurations of zero temperature scalar particles (bosons). The boson field
is described by a complex wave function $\Psi$ and the matter
Lagrangian possesses a $U(1)$ internal symmetry. This symmetry leads to
the existence of a conserved charge $N$ which is interpreted as the
total number of bosons. The first boson star solutions were found by
Kaup \cite{kaup1} and independently by Ruffini \& Bonazzola
\cite{RB}. These authors examined spherically symmetric boson stars in
General Relativity (GR) for which the only self interaction 
term in the matter Lagrangian is the
boson mass. They found that the solutions exhibit qualitatively similar
behaviour to those for neutron stars and white dwarfs: they form a
single parameter family whose ADM mass $M_{ADM}$ and particle number $N$ 
vary smoothly with the parameter and have coinciding maxima and
minima. However, the ADM masses of the stars were found to be extremely
small, of order $M_{pl}^{2}/\alpha$, where $M_{pl}:=\sqrt{\hbar/G}$ is
the Planck mass and $\alpha$ is the mass of the bosons. Colpi, Shapiro
and Wasserman \cite{colpi} studied GR boson stars whose matter
Lagrangian includes a quartic self interaction term. They found that this term
dominates the pressure of the stars and increases the ADM mass to
order $M_{pl}^{3}/\alpha^{2}$.

The stability of boson stars in GR has
been studied by, amongst others, Kusmartsev, Mielke \& Schunck
\cite{kusmar1}-\cite{kusmar2}, who used an approach based 
on catastrophe theory.
To use this method one must identify two conserved charges that vary
smoothly with some parameter which labels the solutions and whose
extrema occur at the same values of this parameter. Then a plot of
one charge against the other will show cusps at the extremal points
and the first of these cusps marks the onset of dynamical
instability. For GR boson stars, the appropriate conserved charges are
$N$ and $M_{ADM}$, and Jetzer \cite{jetzer} has proved that these two
quantities do have coinciding extrema both for pure boson stars and
for mixed boson-fermion stars.

There has been renewed interest recently in generalising GR by adding
one or more scalar (dilaton) gravitational fields. These extra
fields appear in the low energy
limits of super string and super-gravity theories. The simplest
low energy string theory effective action is approximated by the
$\omega=-1$ action of Brans Dicke (BD) theory \cite{BD}, which includes one
dilaton field. This choice of BD parameter has
been ruled out by solar system experiments that imply that
$\omega\ge 500$. However, these experiments do not place such strong
limits on more general
Scalar Tensor (ST) theories in which the coupling parameter $\omega$
is allowed to vary. Thus it is conceivable that the gravitational
interaction can be described by a ST theory in which $\omega\approx
-1$ at some early time when string theory is still valid, and that 
this parameter has increased to it present value as the Universe has
evolved to its present state. 

In the original (Jordan frame) formulation of BD theory, Lee
\cite{lee1} has shown that $M_{ADM}$ is not a good description of the
total energy of a spacetime. In particular, it is not conserved by an
isolated source that emits gravitational radiation. To obtain a conserved
mass on must modify $M_{ADM}$ by adding another term that involves the
derivative of the dilaton field.

Several authors have studied boson stars within the framework of ST
gravity. Gunderson \& Jensen \cite{gund} considered boson stars in BD
theory, focusing on $\omega=6$ solutions, and they showed that these
solutions are qualitatively similar to those in GR.
Tao \& Xue \cite{tao} discuss solutions to a scale invariant
variation of BD theory in which the mass term in the matter Lagrangian
couples to the curvature via the dilaton. This theory necessarily
violates the Weak Equivalence Principle and the authors found a
conserved charge associated with both the boson and dilaton
fields. Torres \cite{torres1} examined boson stars in more general ST
theories while Comer \& Shinkai \cite{comer} used catastrophe theory
to examine the stability of ST boson stars.

When one considers a ST boson star in a cosmological setting, the
evolution of the background (cosmological) value of the dilaton
field may differ from the evolution of the dilaton field  near to the
centre of the boson star. This phenomena was originally described by
Barrow as ``gravitational memory'' \cite{barrow}, although he considered
the effect of cosmological evolution on the horizon area of a black
hole. However, as pointed out by Torres, Liddle \& Schunck
\cite{torres2}, ST boson stars also offer a means of testing
theoretically the degree to which the gravitational memory hypothesis
is true, since boson stars are also highly relativistic objects and
have the added advantage of being singularity free. 
These authors have also studied the gravitational evolution of a BD boson
star in a cosmological background \cite{TSL} and they found that a star
that is stable at the current epoch remains stable as one traces its
evolution backwards in cosmological time.

In this paper we examine how one defines
mass in ST gravity and use boson star solutions to illustrate the
properties of three mass definitions: the Schwarzschild mass (which
gives the ADM mass in the asymptotic limit), the Keplerian mass (that
quantity that determines the geodesic motion of non-self gravitating
test particles) and the Tensor mass (the conserved quantity identified
by Lee \cite{lee1}). For the sake of simplicity we
consider only BD theory. In the spirit of the original formulation of
the theory we assume that the Jordan frame is the physical
frame. We focus on $\omega=-1$ strong field solutions since in this
case the field equations are sufficiently different to the GR
equations for strong field scalar-tensor effects to appear. Also, as
mentioned above, this value of the coupling parameter may be relevant
to the study of boson stars in the early Universe. However,  our
analytical results are valid for any value of $\omega$.

The plan of the paper is as follows. In Section \ref{maineqns} we give
the field equations and discuss the boundary conditions that must be 
imposed to obtain boson star solutions. In
Section \ref{mass} we discuss three definitions of quasi-local mass adapted to
the symmetries of the solutions and give an expression for a
generalisation of the ADM mass that is valid in BD theory. 
In Section \ref{asymptotics}
we examine the asymptotic form of the solutions and show that in the
asymptotic limit, the three quasi-local masses each tend towards one
of the generalised ADM masses. In Section \ref{rescaling} we describe how the
physical characteristics of the solutions change under a rescaling of
the dilaton field. In Section \ref{weakfield} 
we discus the weak field limit of the
solutions and derive an expression for the fractional binding energy
in this limit. In Section \ref{strongfield} 
we describe some of the properties of the
strong field solutions, focusing on the $\omega=-1$ coupling and 
highlighting the differences
between the three mass values we are considering. In Section \ref{extremal}
we prove that for all $\omega$, solutions of extremal particle
number are those of extremal Tensor mass. Finally, in Section
\ref{concl} we give some concluding remarks. Throughout this paper 
we use the sign and index conventions of \cite{MTW}. 

\section{Structure Equations and Boundary Conditions}\label{maineqns}

We take as our starting point the action for Brans-Dicke gravity minimally
coupled to a complex boson field $\Psi$:
\be
\label{action}
   S=\int d^{4}x\sqrt{-g}\left[\frac{e^{-\phi}}{16\pi G}
   \left({\cal R}-\omega\nabla^{\sigma}\phi
   \nabla_{\sigma}\phi\right)-\frac{1}{4}\left(\nabla^{\sigma}\Psi^{*}
   \nabla_{\sigma}\Psi+\nabla^{\sigma}\Psi\nabla_{\sigma}\Psi^{*}\right)
   -\frac{\alpha^{2}}{2\hbar^{2}}\Psi^{*}\Psi \right]
\ee
where $\alpha$ is the mass of the boson particles and we have chosen units in 
which $c=1$. The dilaton field $\phi$ is dimensionless and  the
product $e^{\phi}G$ measures the strength of the gravitational coupling.
Varying the metric leads to the field equations
\be
\label{feqns}
   G_{\mu\nu}=(1+\omega)\nabla_{\mu}\phi\nabla_{\nu}\phi
   -(1+\omega/2)g_{\mu\nu}\nabla^{\sigma}\phi\nabla_{\sigma}\phi
   -\nabla_{\mu}\nabla_{\nu}\phi+g_{\mu\nu}\Box\phi
   +8\pi G e^{\phi}T_{\mu\nu}
\ee
where
\be
\label{Tboson}
   T_{\mu\nu}=\frac{1}{2}\left(\nabla_{\mu}\Psi^{*}
   \nabla_{\nu}\Psi+\nabla_{\mu}\Psi\nabla_{\nu}\Psi^{*}
   -g_{\mu\nu}\frac{\alpha^{2}}{\hbar^{2}}\Psi^{*}\Psi\right)
   -\frac{1}{4}g_{\mu\nu}
   \left(\nabla^{\sigma}\Psi^{*}\nabla_{\sigma}\Psi
   +\nabla^{\sigma}\Psi\nabla_{\sigma}\Psi^{*}\right)
\ee
is the energy momentum tensor of the boson field.
Varying the dilaton and boson fields gives the dilaton wave equation
\be
\label{dwave1}
   2\omega\Box\phi={\cal R}+\omega\nabla^{\sigma}\phi\nabla _{\sigma}\phi
\ee
and the boson wave equations
\be
\label{swave1}
   \Box\Psi=\frac{\alpha^{2}}{\hbar^{2}}\Psi,\;\;\;\;
   \Box\Psi^{*}=\frac{\alpha^{2}}{\hbar^{2}}\Psi^{*}.
\ee
The action possesses a global $U(1)$ symmetry which implies
the existence of a conserved current 
\be
\label{current}
   J^{\mu}=\frac{i}{2}\,
   g^{\mu\nu}\left(\Psi^{*}\nabla_{\nu}\Psi
   -\Psi\nabla_{\nu}\Psi^{*}\right),\;\;\;\;\nabla_{\mu}J^{\mu}=0.
\ee
Note that $J^{\mu}$ has no explicit dependence on $\phi$: eqn (\ref{current})
is identical to the GR definition since the boson field is minimally
coupled to ${\cal R}$. For an asymptotically flat spacetime and time-like
current, eqn (\ref{current}) gives a conserved (time-independent) charge
\be
\label{ndens}
   N=\int d^{3}x\sqrt{h}\,n_{\mu}J^{\mu}
\ee
where the integral is taken over an arbitrary space-like hypersurface
with time-like unit normal $n_{\mu}$ and induced metric $h_{\mu\nu}=
g_{\mu\nu}+n_{\mu}n_{\nu}$. 
In contrast with the solutions discussed in \cite{tao}, there in no
conserved charge associated with the dilaton field. Although we have written
the dilaton terms on the right of eqn (\ref{feqns}), so that they contribute
to the total energy momentum tensor,  we interpret $\phi$
as an extra component of the gravitational field, not as an additional
particle field.

We consider static, spherically symmetric equilibrium solutions.
Using the standard orthogonal $\{t,r,\theta,\varphi\}$ coordinate
basis we write the line element as 
\be
\label{lineelement}
   ds^{2}=-e^{2\nu}dt^{2}+e^{2\lambda}dR^{2}+R^{2}d\theta^{2}
   +R^{2}\sin^{2}\theta \, d\varphi^{2}
\ee
where $\nu$ and $\lambda$ are functions of $R$. We denote the normalised 
time-like Killing vector field by $\xi^{\mu}=(e^{-\nu},0,0,0)$.
Our solutions will be singularity and horizon free, so the coordinate 
system defined by eqn (\ref{lineelement}) is well behaved. We look for
solutions of minimum energy. One can show \cite{FLP} that the form 
of $\Psi$ compatible with this requirement is given by
\be
\label{defboson}
   \Psi=\frac{P}{\sqrt{8\pi
G}}\exp\left(i\frac{\alpha\Omega}{\hbar}t\right),
\ee
where $\Omega$ is a real dimensionless constant and $P$ is a real
dimensionless function of $R$. A further requirement is that the
eigenfunction $P(R)$ possesses no nodes, so that the bosons are in
their ground state. Then the product $\Omega\alpha$ 
defines the ground-state energy of the bosons 
in the zero node eigenstate $P(R)$. From the line element 
(\ref{lineelement}) and the wave-function (\ref{defboson}), 
it is easy to show that $T_{\mu\nu}$ obeys the weak energy condition.

Before writing the equations of motion,
we define a new dimensionless radial coordinate
\be
\label{newrad}
   r:=\frac{\alpha R}{\hbar}.
\ee
Then, using eqns (\ref{lineelement}) and 
(\ref{defboson}), the independent components of the field
and wave equations reduce to the following system of coupled ODEs:
\be
\label{Gtt}
  \nu^{\prime}=\frac{1}{2\phi^{\prime}r-4}\left[e^{2\lambda+\phi}P^{2}r\left(
  1-\Omega^{2}e^{-2\nu}\right)-e^{\phi}P^{\prime 2}r
  -\phi^{\prime 2}r\omega-4\phi^{\prime}-\frac{2}{r}\left(e^{2\lambda}-1
  \right)\right],
\ee
\begin{eqnarray}
\label{Grr}
   \lambda^{\prime}=\frac{e^{\phi}r}{4(2\omega+3)}\left[
   e^{2\lambda}P^{2}\left(\Omega^{2}e^{-2\nu}\left(2\omega+5\right)
   +2\omega-1\right)+P^{\prime 2}\left(2\omega+1\right)\right]  \nonumber \\
    +\frac{\phi^{\prime 2}r\omega}{4}+\frac{\phi^{\prime}\nu^{\prime}r}{2}
   -\frac{1}{2r}\left(e^{2\lambda}-1\right),
\end{eqnarray}
\be
\label{dwave2}
   \phi^{\prime\prime}=\frac{e^{\phi}}{2\omega+3}\left[e^{2\lambda}P^{2}
   \left(2-\Omega^{2}e^{-2\nu}\right)+P^{\prime 2}\right]+\phi^{\prime 2}
   +\phi^{\prime}\left(\lambda^{\prime}-\nu^{\prime}-\frac{2}{r}\right)
\ee
and
\be
\label{swave2}
   P^{\prime\prime}=e^{2\lambda}P\left(1-\Omega^{2}e^{-2\nu}\right)
   +P^{\prime}\left(\lambda^{\prime}-\nu^{\prime}-\frac{2}{r}\right),
\ee
where the prime denotes $\frac{d\;}{dr}$. 
To solve these equations we impose the boundary conditions
\be
\label{bdry}
   e^{\lambda_{0}}=1,\;\;P^{\prime}_{0}=0,\;\;\phi^{\prime}_{0}=0,\;\;
   P_{\infty}=0
\ee
where throughout this paper the subscripts `0' and `$\infty$' denote values at
 $r=0$ and $r=\infty$ respectively. The first three conditions enforce 
regularity of the solutions at the origin, while the last 
condition ensures that the boson matter is localised and the solutions
are asymptotically flat. For any given $\omega$ the solutions may be
parameterised by $P_{0}$ and $\phi_{\infty}$.
The field equations then become eigenvalue equations for $\Omega$.
Note that
the value of $e^{\nu}$ at infinity is arbitrary: eqns (\ref{Gtt}) to
(\ref{swave2}) are invariant under the rescaling $\nu\rightarrow\nu+k$, $\Omega
\rightarrow\Omega e^{k}$ where $k$ is constant. We use this freedom
to set 
\be\label{bdry2}
   e^{\nu_{\infty}}=1
\ee
 which fixes the scale of our time coordinate
and consequently our unit of energy.
The field equations automatically lead to the asymptotic
conditions 
\be\label{bdry3}
   P^{\prime}_{\infty}=\nu^{\prime}_{\infty}=\phi^{\prime}_{\infty}=
   \lambda_{\infty}=0.
\ee

From eqns (\ref{current}) and (\ref{defboson}), $J^{\mu}$ is
parallel to $\xi^{\mu}$ and has the non-zero
component
\be
   J^{0}=\frac{\hbar\Omega P^{2}}{8\pi\alpha G}e^{-2\nu}.
\ee
Choosing a hypersurface orthogonal to $\xi^{\mu}$ and combining the
above expression with eqn (\ref{ndens}) we have
\be
\label{ndens2}
   N=\frac{1}{2}\integ e^{\lambda-\nu}
   \Omega P^{2}r^{2}\,dr
\ee
where we have expressed $N$ in units of $M_{pl}^{2}/\alpha
=\hbar/\alpha G$.
Since there are no Maxwell terms present in the action (\ref{action}),
the bosons have no electromagnetic charge and we interpret the quantity
$\frac{1}{\alpha}N$ as the total boson particle number
so that $N$ is the total rest mass of the star. This latter quantity
is simply the sum of the masses of the bosons that make up the star,
as measured by some non-gravitational experiment. From the rest mass
we define the Newtonian mass
\be\label{MNdef}
   M_{N}:=Ne^{\phi_{\infty}}
\ee
which measures the gravitational mass (or energy) of a star whose
bosons are dispersed to infinity. 

Finally we note that to recover the GR equations of motion and their 
corresponding solutions we take the limit $\omega\rightarrow\infty$
which implies that $\phi\rightarrow\phi_{GR}$, where $\phi_{GR}$  is constant. 
Since we have included Newton's constant
$G$ in the action (\ref{action}), we have used up the freedom to
scale the dilaton (which is equivalent to rescaling the unit of mass)
and so we have the additional requirement $\phi_{GR}=0$. 

\section{Quasi-Local and ADM Masses}\label{mass}

Operationally, in Newtonian theory one determines the active gravitational
mass $GM$ of a gravitating source by applying
Kepler's third law. A test particle in a
circular orbit of radius $R$ and angular velocity $d\varphi/dt$ about
the source measures an active mass
\be
\label{keplermass}
   GM=\;\lim_{R\to\infty}
   \left[R^{3}\left(\frac{d\varphi}{dt}\right)^{2}\right].
\ee
This mass is a product of the gravitational coupling strength $G$ and
the rest mass $M$ of the source. If the source is spherically 
symmetric, eqn (\ref{keplermass}) is valid for all $R$ and
it measures the gravitational mass $GM(R)$ contained
within the 2-sphere $\Sigma(R)$ of radius $R$.

For metric theories of gravity the determination of mass is more difficult.
All we can discuss is some total gravitational mass,
which is determined by the metric $g_{\mu\nu}$ and any 
additional tensor or scalar gravitational fields appearing in 
the theory. Furthermore, for metric
theories other than GR, the Strong Equivalence Principle (SEP, defined in 
\cite{will}) is violated. Because of this, the mass we measure using Kepler's
third law depends upon the test particle's gravitational binding energy.
Hence, in an alternative theory of gravity we cannot in general separate the
gravitational coupling strength from the rest mass of the source.

Returning to the boson star solutions, we first discuss three definitions 
of quasi-local mass (QLM), adapted to the symmetries of the line
element (\ref{lineelement}). As is well known, QLM definitions are of limited
utility. One reason for this is that it is difficult to write a 
coordinate invariant conservation law for any quasi-local
quantity. However, in a static spacetime, the time-like Killing
vector field $\xi^{\mu}$ provides a natural splitting of the
spacetime into a set of space-like hypersurfaces of constant time on
which one can formulate an expression for the QLM. The existence of
$\xi^{\mu}$ guarantees that quantities defined on the constant time
slices are conserved. There are still an infinite number of ways of
defining QLM in a static spacetime. However, an acceptable definition of
the mass $M({\cal U})$ of a sub-manifold ${\cal U}$
must satisfy the following three requirements: (a) $M$ is non-negative,
(b) $M=0$ everywhere in  Minkowski spacetime, and (c)
given two sub-manifolds ${\cal U}$ and ${\cal V}$, such that
${\cal U} \subset {\cal V}$, then $M({\cal U})\leq M({\cal V})$. We
also feel that any global mass defined as the limit of some QLM is
only valid if the QLM satisfies these conditions, and we use this criterion
to select physically reasonable definitions of
mass for the static boson star solutions we consider here.

We first consider the generalised Schwarzschild mass, which is given
by any one of the familiar definitions
\be
\label{mass1}
   m_{S}(r)=\frac{r}{2}\left(1-e^{-2\lambda}\right)=\int_{0}^{r}
   \frac{\tr ^{2}}{2}\rho\,d\tr
   =\int_{0}^{r}\left[\frac{1}{2}\left(1-e^{-2\lambda}\right)
   +\lambda^{\prime}\tr e^{-2\lambda}\right]\;d\tr
\ee
where $m_{S}(r)$ is measured in units of $M_{pl}^{2}/\alpha$
and $\rho$ is the total energy density defined by
\be
\label{rho}
   \rho:=G_{\mu\nu}\xi^{\mu}\xi^{\nu}.
\ee
Equation (\ref{mass1}) is the definition of mass adopted in all references on 
boson stars cited in the introduction except \cite{comer}, and it gives an 
unambiguous definition of the energy of a spherically symmetric,
asymptotically flat
system in GR (see \cite{MTW}, p.\  603): provided $\rho$ is
non-negative, $m_{S}$ satisfies conditions (a) to (c) outlined above
and in the limit $r\rightarrow\infty$, $m_{S}$ tends to the ADM mass. However,
in Brans Dicke theory one or more of these conditions may be
violated even for a physically reasonable energy momentum tensor.
From eqns (\ref{feqns}) and (\ref{rho}) the total energy density of the 
boson star is given by
\be\label{edensity}
   \rho=e^{-2\nu}G_{00}=\left(1+\frac{\omega}{2}\right)e^{-2\lambda}
   \phi^{\prime 2}-\Delta\phi+\frac{e^{\phi}}{2}\left(\Omega^{2}P^{2}
   e^{-2\nu}+P^{2}+P^{\prime 2}e^{-2\lambda}\right)
\ee
where
\be\label{3boxs}
   \Delta\phi:=\frac{e^{-\lambda}}{r^2}\frac{d\;}{dr}
   \left(r^{2}e^{-\lambda}\phi^{\prime}\right).
\ee
This expression contains non-positive definite terms so although $T_{\mu\nu}$ 
obeys the weak energy condition the total density may still be
negative in some regions of a solution,
leading to the condition $m_{S}^{\prime}<0$ (in
violation of requirement (c)). This is indeed the case for the
$\omega=-1$ solutions discussed later, and there even exist weak field
solutions for which $m_{S}<0$ everywhere. The fact that the QLM
$m_{S}$ may have a negative gradient leads one to conclude that the
ADM mass, which as we shall see below is given by
$\lim_{R\to\infty}m_{S}$, fails to account for all of the energy of
the spacetime.

As an alternative to eqn (\ref{mass1}), we may determine the gravitational
mass enclosed within a two sphere $\Sigma(r)$ by measuring the analogue of the
gravitational acceleration of a test particle at radius $r$. More specifically,
we consider circular geodesic orbits and use eqn (\ref{keplermass}) to
determine the mass enclosed by the orbit. This definition of mass is valid
in a static, spherically symmetric spacetime. In this case we may project a
particle orbit of constant $r$ onto a hypersurface orthogonal to $\xi^{\mu}$
to form a circle about the centre of symmetry. The set of all orbits at
this radius, projected in the same way, then form a closed two sphere 
$\Sigma(r)$. The matter and gravitational fields contained within this
sphere remain constant in time, and so the orbital mass is well defined.
Since the SEP is violated in BD theory, the value of the orbital mass we
measure depends upon the gravitational
binding energy of the orbiting test particle.

We consider two orbital masses: the Keplerian and the Tensor mass.
A {\em non-self gravitating} test particle in a circular geodesic orbit
in the geometry of eqn (\ref{lineelement}) moves
with an angular velocity
\be
\label{angvel}
   \frac{d\varphi}{dt}=e^{\nu}\sqrt{\frac{\nu^{\prime}}{r}}
\ee
as measured by an observer at infinity.
Use of eqn (\ref{keplermass}) then leads to the Keplerian mass function
\be
\label{mass2}
   m_{K}(r)=\nu^{\prime}r^{2}e^{2\nu}
\ee
in units of $M_{pl}^{2}/\alpha$. The Tensor mass we define similarly
by considering the circular orbit of a test particle in the Einstein frame,
whose metric ${\tilde g_{\mu\nu}}$ is conformally related to the 
physical (Jordan frame) metric $g_{\mu\nu}$ by the transformation
${\tilde g_{\mu\nu}}=e^{-\phi}g_{\mu\nu}$. The orbital angular velocity 
in the Einstein frame is given by
\be
\label{Eangvel}
   \frac{d\varphi}{dt}=e^{\nu}\sqrt{\frac{2\nu^{\prime}-\phi^{\prime}}
   {r(2-r\phi^{\prime})}}
\ee
Kepler's third law (\ref{keplermass}) then gives the Tensor
mass function
\be
\label{mass3}
   m_{T}(r)=\frac{r^{2}e^{2\nu}(2\nu^{\prime}-\phi^{\prime})}
   {2-\phi^{\prime}r},
\ee
which may be interpreted as the mass measured by an orbiting test
black hole in the Jordan frame \cite{hawking}. As with the previous
mass functions, the Tensor mass is given in units of $M_{pl}^{2}/
\alpha$. Note that the conformal transformation used to define the
Tensor mass is merely a computational device. We use it to
find a frame in which the test particle's motion {\em is} geodesic
and use this motion to define the Tensor mass. We could equally
well define a different orbital mass by transforming to a different
conformal frame.

Numerical calculations for the $\omega=-1$ BD solutions
show that both $m_{K}$ and $m_{T}$ are non-decreasing 
functions of $r$. From eqns (\ref{angvel}) and (\ref{Eangvel}), both mass 
functions are also non-negative definite (as a corollary
eqn (\ref{mass2}) implies that $e^{\nu}$ is a non-decreasing 
function of $r$). In Minkowski spacetime one has 
$m_{K}=m_{T}=0$ everywhere. Hence the Keplerian and 
Tensor masses satisfy conditions (a) and (b) given above and, at least
for the $\omega=-1$ solutions studied later, they appear to satisfy
condition (c).

Lee \cite{lee1} has shown that one can derive ADM-like masses in 
Brans-Dicke gravity from the superpotential
\be
\label{spotential}
   H^{\mu[\nu\alpha]\beta}=(-g)e^{n(\phi_{\infty}-\phi)}\left(
   g^{\mu\nu}g^{\alpha\beta}-g^{\mu\alpha}g^{\nu\beta}\right)
\ee
where $n$ is usually taken to be an integer. Note that this expression
is not unique due to the ambiguity in the choice of pseudo-tensor for
the gravitational field. Furthermore, the index $n$ is undetermined,
which reflects the additional ambiguity we have in determining the
contribution the dilaton field makes to the total gravitational energy
momentum.

The potential satisfies the conservation law $H^{\mu\nu\alpha\beta}
_{\;\;\;\;\;\;\;\;,\nu\alpha\beta}=0$ and on integrating this expression
and using Stoke's theorem twice we have the generalised ADM mass
\be
\label{mADM1}
   M_{n}=-\frac{1}{16\pi}\int\left[(-g)e^{n(\phi_{\infty}-\phi)}(g^{00}g^{ij}
   -g^{0i}g^{0j})\right]_{,j} \,d^{2}\Sigma_{i}
\ee
where the integral is performed over the 2-sphere
$\Sigma(r)$ in the limit $r\rightarrow\infty$ and the metric must be
expressed in rectangular (Minkowski) coordinates. 
Our definition of $H^{\mu\nu\alpha\beta}$
differs from the one given in \cite{lee1} by a factor of $e^{n\phi_{\infty}}$
which is included here so that $M_{n}$ is expressed in units 
of $M_{pl}^{2}/\alpha$.
 For the metric defined by eqn (\ref{lineelement}),
the integral (\ref{mADM1}) evaluates to
\be
\label{mADM2}
   M_{n}=\;\lim_{r\to\infty}
   \left[r\left(1-e^{-\lambda}\right)\right]+\frac{n\phi_{1}}{4}
\ee
where 
\be\label{Sdef1}
   \phi_{1}=\lim_{r\to\infty}(r^{2}\phi^{\prime}).
\ee
In the following Section we shall relate this 
expression to the asymptotic limits of the quasi-local masses.

\section{The Asymptotic Form of the Solutions}\label{asymptotics}

Expanding the metric and mass functions in powers of $1/r$ about $r=\infty$,
one can show that to order $1/r$ the line element is given by 
\be
\label{lineelement2}
   ds^{2}=- \left( 1-\frac{2M_{K}}{r} \right) dt^{2}+
   \left(1+\frac{2(M_{K}-\phi_{1})}{r}\right) dr^{2}
   +r^{2}d\theta^{2}+r^{2}\sin^{2}\theta \, d\varphi^{2}.
\ee
where $M_{K}$ is the limit of eqn ({\ref{mass2}) as $r\rightarrow
\infty$ and $\phi_{1}$ is defined in eqn (\ref{Sdef1}). Since
$\lim_{r\to 0}(r^{2}\phi^{\prime})=0$ this quantity may also be
written as
\be\label{Sdef2}
   \phi_{1}=\integ \frac{d\;}{dr}
   \left(r^{2}\frac{d\phi}{dr}\right)\;dr.
\ee

Equation (\ref{lineelement2}) is a special case of the vacuum BD 
solution given in \cite{BD} and the asymptotic solution is  determined 
by two parameters which we take here to be
$\phi_{1}$ and $M_{K}$. For this form of the metric,
 the boson wave equation (\ref{swave2}) has the asymptotic solution
\be
\label{Papprox}
   P= r^{b}e^{-\kappa r}\left[1+{\cal O}\left(\frac{1}{r}\right)\right]
\ee
where
\be
\label{pcoeffs}
   b=-1-\kappa(M_{K}-\phi_{1})+\frac{\Omega^{2}}
 {\kappa}M_{K},\;\;\;\;\kappa=\sqrt{1-\Omega^{2}}.
\ee
The boson field falls off exponentially with $r$,
far  more rapidly than any other field. This justifies the use of the 
name ``star'' in describing these solutions: 
although the object has no well defined surface,
the boson matter is still highly localised. The parameter $\kappa$
may be interpreted as the reciprocal radius of the star and must be real
for an acceptable solution. This implies that $\Omega\leq 1$.

These solutions are asymptotically flat so that for large $r$ we have
$1-e^{-\lambda}\sim\frac{1}{2}(1-e^{-2\lambda})$. Hence,
setting $n=0$ in eqn (\ref{mADM2}) and comparing the result with eqn
(\ref{mass1}) gives
\be
\label{mlimit1}
   M_{0}=M_{ADM}=\lim_{r\to\infty}m_{S}
\ee
and we see that, as mentioned above,
the generalised Schwarzschild mass tends to the ADM mass
in the Jordan frame. Taking the leading terms in the expansion of the 
Keplerian and Tensor masses, one can show that
\be
\label{mlimit2}
   M_{K}=M_{4}=M_{ADM}+\phi_{1},\;\;\;\;\;\;
   M_{T}=M_{2}=M_{ADM}+\frac{\phi_{1}}{2}.
\ee
Note that $M_{T}$ is the ADM mass of the star in the
Einstein frame, which is not equal to $M_{ADM}$ since the derivative 
of the conformal factor $e^{-\phi}$ which relates the two frames 
is non-vanishing. Comer and Shinkai \cite{comer} use the correct
definition of mass (the Schwarzschild mass in the Einstein frame)
but, as we have seen, this quantity is not the same as the Jordan
frame ADM mass (this contradicts a statement given in \cite{comer}).
Numerical studies show that $\phi^{\prime}>0$ for all $\omega=-1$
solutions investigated. This implies that
\be
\label{mrelations}
   M_{ADM}\leq M_{T}\leq M_{K}
\ee
for these solutions.

For each value of the mass $M_{n}$ there is an associated fractional
binding energy $E_{n}$ per unit boson  mass given by
\be
\label{benergy}
   E_{n}=(M_{n}-M_{N})/M_{N}.
\ee
This expression is independent of the unit of mass chosen. The choice
of which $E_{n}$ correctly measures the binding energy of the star depends 
upon which mass value $M_{n}$ corresponds to the true energy of the star.

\section{Dilaton Rescaling}\label{rescaling}

As mentioned above, the solutions may be parameterised by the pair
$P_{0},\phi_{\infty}$. 
The former quantity determines the energy density of the boson 
field at the origin and for each pair of parameters 
$(P_{0}, \phi_{\infty})$ there is a unique value of pair of $\Omega 
e^{-\nu_{0}}$ that gives a zero node solution with the 
boundary conditions (\ref{bdry}) and (\ref{bdry2}). The 
product $\Omega e^{-\nu_{0}}$ 
measures the bosons' ground state energy per unit boson
mass at the origin, as seen by an observer at $r=\infty$. This quantity
increases monotonically with $P_{0}$ and the pair $(\Omega e^{-\nu_{0}},
\phi_{\infty})$ serves as an alternative parameterisation of the solutions.
Note that $\Omega e^{-\nu_{0}}$ is invariant under the rescaling of
$e^{\nu}$ described before eqn (\ref{bdry2}).

The equations of motion and the mass equations (\ref{MNdef}), 
(\ref{mass1}), (\ref{mass2}), (\ref{mass3}) and (\ref{mADM2}) 
are invariant under the rescaling 
\be
\label{rescale}
e^{\phi}\rightarrow k^{2}e^{\phi},\;\;\;\;\;\;P\rightarrow \frac{P}{k}
\ee
where $k$ is a constant, while $N$ re-scales under eqn 
(\ref{rescale}) as
\be
N\rightarrow\frac{N}{k^{2}}.
\ee
This rescaling swaps energy between the dilaton part of the
gravitational field and the boson energy density is such a way as to 
leave the total gravitating energy invariant. Given a set of solutions
\be
   {\cal S}(P_{0};\phi_{\infty})=\left\{M(P_{0};\phi_{\infty}),\;
   N(P_{0};\phi_{\infty})\right\}
\ee
parameterised by $P_{0}$ for some fixed $\phi_{\infty}$, where $M$ is
any asymptotic or quasi-local mass value,
eqn (\ref{rescale}) generates a new physically distinct set
\be
\label{newsoln}
   {\cal S}(P_{0};\phi_{\infty}+2\log k)=\left\{M(kP_{0}
   ;\phi_{\infty}),\;\frac{1}{k^{2}}N(kP_{0};\phi_{\infty})\right\}.
\ee
This new set of solutions have exactly the same mass and binding energy as
the first set, but consists of stars of different total particle
numbers with compensating changes in their gravitational coupling strength.  
Hence, eqn (\ref{newsoln}) may be used to compare boson star solutions 
in cosmological settings with different values of $\phi_{\infty}$.

\section{Weak Field Solutions}\label{weakfield}

In this section we consider the behaviour of the zero node solutions in 
the weak field limit, which we define
as the limit in which $P(r)$ is small but non-zero.
Following Kaup \cite{kaup2}, we write the boson field amplitude
as
\be
   P=\ve\Pi(a)
\ee
where $0<\ve\ll 1$ and $a$ is a new radial coordinate defined
by $a=\sqrt{\ve}r$. The function $\Pi$ is defined to have the boundary
value $\Pi_{0}=1$, so that $\ve$ measures the degree to which the
solutions differ from flat spacetime. We use $\ve$ as an 
expansion parameter and to first order in $\ve$ we expand the
metric functions, dilaton field and energy eigenvalue about flat
spacetime as
\be\label{wfvardef}
   \lambda=\frac{\ve}{2}A(a)\;\;\;\;\nu=\frac{\ve}{2}B(a) 
   \;\;\;\;\phi=\phi_{\infty}+\ve\Phi(a)\;\;\;\;
   \Omega=1-\frac{\ve}{2}\Gamma
\ee
where $\Gamma(P_{0},\phi_{\infty})$ is positive and constant for each solution.
Equations (\ref{Gtt}) to (\ref{swave2}) then reduce to
\be
\label{linGtt}
   \left(a^{2}B^{\prime}\right)^{\prime}=
   \frac{2(\omega+2)}{2\omega+3}a^{2}\Pi^{2}e^{\phi_{\infty}}
   +{\cal O}(\ve),
\ee
\be
\label{linGrr}
   (aA)^{\prime}=\frac{2(\omega+1)}{2\omega+3}
   a^{2}\Pi^{2}e^{\phi_{\infty}}
   +{\cal O}(\ve),
\ee
\be
\label{lindwave}
   \left(a^{2}\Phi^{\prime}\right)^{\prime}=\frac{a^{2}\Pi^{2}
   e^{\phi_{\infty}}}{2\omega+3}
   +{\cal O}(\ve)
\ee
and
\be
\label{linswave}
   \left(a^{2}\Pi^{\prime}\right)^{\prime}=a^{2}\Pi(B+\Gamma)
   +{\cal O}(\ve)
\ee
where throughout this section a prime denotes $\frac{d\;}{da}$.

Using eqns (\ref{ndens2}), (\ref{MNdef}), (\ref{mass1}) 
and (\ref{edensity}), the derivatives 
of eqns (\ref{mass2}) and (\ref{mass3}),
and the weak field equations of motion the quantities $M_{N}$,
$M_{ADM}$, $M_{T}$ and $M_{K}$ may be written as the integrals
\be
\label{linnum}
   M_{N}=e^{\phi_{\infty}}N=\sqrt{\ve}\integ 
   \frac{a^{2}\Pi^{2}e^{\phi_{\infty}}}{2}\,da+{\cal O}(\ve^{\thrhalf})
\ee
\be
\label{linmass1}
   M_{ADM}=M_{0}=\sqrt{\ve}\integ a^{2}\Pi^{2}
   e^{\phi_{\infty}}\frac{(\omega+1)}{2\omega+3}
   \,da+{\cal O}(\ve^{\thrhalf})
\ee
\be
\label{linmass3}
   M_{T}=M_{2}=\sqrt{\ve}\integ 
   \frac{a^{2}\Pi^{2}e^{\phi_{\infty}}}{2}\,da+{\cal O}(\ve^{\thrhalf})
\ee
\be
\label{linmass2}
   M_{K}=M_{4}=\sqrt{\ve}\integ a^{2}\Pi^{2}
   e^{\phi_{\infty}}\frac{(\omega+2)}{2\omega+3}\,da
   +{\cal O}(\ve^{\thrhalf}).
\ee
Equation (\ref{linmass1}) shows that, in the weak field limit,
the ADM mass is negative for $\omega<-1-\ve$. This is 
a consequence of treating $\phi$ solely as a matter field. The energy density
may be written partly in terms of the weak field variables to give
\be
\label{lindens}
   \rho=-\frac{\ve^{2}}{a^{2}}\left(a^{2}\Phi^{\prime}\right)^{\prime}
   +8\pi G e^{\phi}T_{\mu\nu}\xi^{\mu}\xi^{\nu},
\ee
where we have shown the dilaton contribution to lowest non-zero
order in $\ve$ and one can show that the second term on the right is
also of order $\ve^{2}$. 
Equation (\ref{lindwave}) implies that $(a^{2}\Phi^{\prime})^{\prime}$
is positive, so that 
the dilaton contribution to the total density in eqn (\ref{lindens}) 
is negative. As $\omega$ decreases below $\omega\sim -1$, the derivatives
of the dilaton field increase in magnitude and the total density 
becomes dominated by the (negative) dilaton term.
This leads to a negative Schwarzschild mass gradient for all $r$ and
so gives a negative quasi-local mass $m_{S}$, and hence negative $M_{ADM}$.
When $\omega>-1+\ve$ the coupling between the dilaton and the
curvature is weak enough for the boson field density to dominate over
the dilaton density, so that in this case both $m_{S}$ and $M_{ADM}$ 
are positive. For $|\omega-1|\sim\ve$ we need to go to higher order 
in $\ve$ to
calculate $M_{ADM}$. However, numerical calculations show that
$M_{ADM}$ increases smoothly from negative to positive values
as $\omega$ increases over the interval $(-1-\ve,-1+\ve)$.

From eqns (\ref{linnum}), (\ref{linmass3}) and (\ref{linmass2}) we 
see that, asymptotically, the weak field solution describes the gravitational
field about a central source of rest mass $N$ so that to lowest order
in $\ve$ the Tensor and Keplerian masses may be written as the products
$M_{2}=G_{2}N$, $M_{4}=G_{4}N$ 
where the coupling strengths $G_{2}$, $G_{4}$ are given by
\be
   G_{2}=e^{\phi_{\infty}}\;\;\;\;\;\;G_{4}=e^{\phi_{\infty}}
   \frac{2(\omega+2)}{2\omega+3}.
\ee
Hence the weak field solutions are Newtonian in the sense that the rest
mass of the star may be separated from the gravitational coupling 
strength. However, this coupling is not unique and its strength
depends upon the properties of the orbiting test particle that we use 
to measure the star's mass. The quantity $G_{4}$, often quoted as the
effective gravitational constant in the weak field limit (see for
example \cite{will}), is the coupling strength one would measure in a
Cavendish experiment using non-self gravitating test
particles. Performing the same experiment using test black holes, one
would measure the coupling strength  $G_{2}$.

We use eqns (\ref{linnum}), (\ref{linmass1}) and (\ref{linmass2}) to
calculate the fractional binding energies associated with the ADM and
Keplerian masses. The result is
\be\label{Benergy1}
   E_{0}=\frac{-1}{2\omega+3}+{\cal O}(\ve),\;\;\;\;\;\;
   E_{4}=\frac{1}{2\omega+3}+{\cal O}(\ve).
\ee
Both of these quantities are independent of $\ve$ in their leading term
which is physically unreasonable: as $\ve\rightarrow 0$ and the solution
approaches flat spacetime, we would expect the fractional binding
energy to vanish as it does in the GR case \cite{kaup2}. The fact that
$E_{0}$ and $E_{4}$ behave in a non-physical way indicates that
neither the ADM mass nor the Keplerian mass correctly describes the total
energy of the star.

As is apparent when considering eqns (\ref{linnum}) and
(\ref{linmass3}), to calculate $E_{2}$ we need to express $M_{N}$ and
$M_{T}$ to order $\ve^{\thrhalf}$. It turns out that the ${\cal
O}(\ve^{\thrhalf})$ terms of both of these quantities depend only upon
the ${\cal O}(\ve)$ weak field variables defined in eqn
(\ref{wfvardef}). One can then show that the fractional binding 
energy associated with the Tensor mass is given by
\be\label{Benergy2}
   E_{2}=-\frac{\ve}{\cal N}\integ 
   \frac{a^{3}\Pi^{2}}{2\omega+3}\left[\Phi^{\prime}(\omega+2)
   +B^{\prime}\frac{(2\omega+1)}{4}\right]\;da
   +{\cal O}(\ve^{2}),
\ee
where we have defined
\be\label{Ndef}
   {\cal N}:=\integ \frac{a^{2}\Pi^{2}}{2}\;da
\ee
so that $\sqrt{\ve}{\cal N}$ is the rest mass to lowest order in $\ve$. 
We need to simplify eqn (\ref{Benergy2}). Integrating eqns (\ref{linGtt}) and
(\ref{lindwave}) one can show that
\be
   \Phi^{\prime}=\frac{B^{\prime}}{2(\omega+2)}
\ee
and substituting this result into eqn (\ref{Benergy2}) gives
\be\label{Benergy3}
   E_{2}=-\frac{\ve}{\cal N}\integ 
   \frac{a^{3}\Pi^{2}}{4}B^{\prime}\;da
   +{\cal O}(\ve^{2}).
\ee
Integrating both $a^{2}\Pi^{\prime 2}$ and
$\left[a\Pi^{\prime}\left(a^{2}\Pi^{\prime}\right)^{\prime}\right]$ by
parts and using eqn (\ref{linswave}) one can show that 
\be\label{dB1}
   \integ a^{3}\Pi^{2}B^{\prime}\;da
   =-2\integ a^{2}\Pi^{2}(\Gamma+B)\;da.
\ee
Similarly, integrating both 
$a^{2}B^{\prime 2}$ and
$\left[aB^{\prime}\left(a^{2}B^{\prime}\right)^{\prime}\right]$ by
parts and using eqn (\ref{linGtt}) one can show that 
\be\label{dB2}
   \integ a^{2}\Pi^{2}B\;da
   =-2\integ a^{3}\Pi^{2}B^{\prime}\;da.
\ee
Substituting eqn (\ref{dB2}) into (\ref{dB1}) gives
\be
   \integ a^{3}\Pi^{2}B^{\prime}\;da
   =\frac{2\Gamma}{3}\integ a^{2}\Pi^{2}\;da
   =\frac{2\Gamma}{3}{\cal N}
\ee
where we have used eqn (\ref{Ndef}) to obtain the last
equality. Combining the above result with eqn (\ref{Benergy3}) gives
\be\label{Benergy4}
   E_{2}=-\frac{\ve\Gamma}{6}+{\cal O}\left(\ve^{2}\right).
\ee
This is identical to the expression for the GR binding energy found by
Kaup \cite{kaup2} and, in particular, is independent of
$\omega$ and vanishes as $\ve\rightarrow 0$. Hence all weak field BD
boson stars are energetically stable.

\section{Strong Field Solutions}\label{strongfield}

We have numerically integrated the field equations (\ref{Gtt}) 
to (\ref{swave2}) for the case $\omega=-1$ over the parameter 
range $0\leq P_{0}\leq 1.6$. Results of the integration are given in
Figure 1, which shows the behaviour of the three asymptotic masses we
have discussed and how they relate to the Newtonian mass $M_{N}$. We
have chosen the boundary value $e^{\phi_{\infty}}=1$ so that $M_{N}$ is
numerically equal to $N$ and our results may may be compared with
those of other authors, although this choice is not physically
realistic: by setting $\omega=-1$ we are implicitly describing boson
star solutions in the early (string-theory) Universe. In general, ST
cosmological solutions show that the cosmological value of $\phi$
decreases with cosmic time, so that we should choose the boundary
value $e^{\phi_{\infty}}\gg 1$. However, from the scaling relation
(\ref{rescale}), changing the asymptotic value of $\phi$ merely
changes the scale of the the horizontal axis of the figure (although
for $e^{\phi_{\infty}}\neq 1$, $M_{N}$ and $N$ are no longer equal). Note
also that the fractional binding energy of any solution is invariant
under the rescaling of $\phi_{\infty}$ (this result is true for all
values of $\omega$). 

The four mass curves exhibit qualitatively similar behaviour to the
ADM mass and particle number curves found for GR boson stars.
Each varies smoothly with $P_{0}$ and each mass
curve reaches an initial maximum before falling off with increasing
$P_{0}$ and tending towards low amplitude oscillations about some low
mass value. There are, however, several quantitative differences
between the GR and BD solutions.

All of the solutions investigated possess a region in which
$-\phi^{\prime\prime}$ is large and negative and dominates over the
boson field contribution to the total energy density. This implies
that $\rho<0$ in this region, which leads to the condition
$m_{S}^{\prime}<0$ there, in violation of condition (c) given in
Section \ref{mass}. The quasi-local Tensor and Keplerian masses $m_{T}$ and
$m_{K}$ were found to be increasing functions of $r$ for all
solutions. 

An obvious feature of the Figure is that the asymptotic masses
$M_{ADM}$, $M_{K}$ and $M_{T}$ differ remarkably from each other. This
is a result of choosing such a low value of $\omega$, which means
that the gradient of $\phi$ is large, even in the asymptotic limit. 
Consequently, for these solutions $\phi_{1}$ is of the same order of
magnitude as $M_{ADM}$ and from eqn (\ref{mlimit2}) the difference
between the values of the three masses at each $P_{0}$ is large.
As $\omega$ increases, the dilaton field becomes less strongly coupled
to the curvature and the ratio $\phi_{1}/M_{ADM}$ decreases for a
given $N$, which implies that both $M_{K}$ and $M_{T}$ tend towards
$M_{ADM}$. From eqns (\ref{linmass1}) and (\ref{linmass3}), for a
given $P_{0}$ the fractional difference between $M_{ADM}$ and $M_{T}$
in the weak field limit is given by
\be
   \frac{M_{T}-M_{ADM}}{M_{T}}=\frac{1}{2\omega+3}.
\ee
Further numerical work suggests that this relation is approximately
true for strong field solutions, with an accuracy that increases with
$\omega$. For the $\omega=6$ coupling chosen by Gunderson \& Jensen
\cite{gund}, the fractional difference between $M_{ADM}$ and $M_{T}$
is around 1/15, which implies that the boson star masses that they
quote are about 93\% of their true values. For the choice $\omega=500$
implied by the current observational constraints, the fractional
difference between the two masses is less that $10^{-3}$ which, as
pointed out in \cite{TSL}, is negligible.

A second obvious feature of the Figure is that the 
$M_{ADM}$, $M_{K}$ and $M_{T}$
curves have extrema located at different values of $P_{0}$. This
feature persists for all finite $\omega$ and is only absent in the
limit $\omega\rightarrow\infty$ where the three masses are identical.
However, it does appear from the figure that the extrema of $M_{T}$
and $M_{N}$ do occur at the same values of $P_{0}$, which suggests
that $M_{T}$ is the correct definition of the total energy of the
star. In the following Section we give a proof that this is indeed 
the case for all values of $\omega$.

\section{Coincidence of Extremal Tensor Mass and Extremal Particle Number  
Solutions}\label{extremal}

We consider variations $\delta M_{T}$ and $\delta M_{N}$ in the Tensor
and Newtonian masses induced by the parameter change $P_{0}\rightarrow
P_{0}+\delta P_{0}$ with
$\phi_{\infty}$ held fixed. Since each pair $P_{0},\phi_{\infty}$
gives a unique asymptotically flat zero node solution to the equations
of structure (\ref{Gtt}) to (\ref{swave2}), we may equivalently
consider arbitrary variations in $M_{T}$ and $M_{N}$ subject to the
constraint that the field equations and the boundary conditions
(\ref{bdry}) to (\ref{bdry3}) hold. 
We shall then end up with a simple relationship between 
$\delta M_{T}$ and $\delta M_{N}$. Note that because of our choice of
coordinates, the variation $\delta$ commutes
with both integration and differentiation with respect to $r$.

We start by taking the variation in the rest mass $N$.
From the definition (\ref{ndens2}) we have
\be\label{varN1}
   \delta N=\integ \frac{r^{2}}{2}
   \left[e^{\lambda-\nu}\Omega P^{2}\delta\lambda
   +e^{\lambda}\delta\left(e^{-\nu}\Omega P^{2}\right)\right]\;dr.
\ee
Using a similar method to the one outlined in \cite{jetzer} we 
rewrite the second term in the above integrand in the following way.
One can easily show that 
\be
   \delta\left(e^{-\nu}\Omega P^{2}\right)=\frac{1}{2\Omega}e^{\nu-\phi}
   \delta\left(e^{\phi-2\nu}\Omega^{2}P^{2}\right)
   -e^{-\nu}\frac{\Omega P^{2}}{2}\delta\phi+e^{-\nu}\Omega P\delta P.
\ee
We solve eqn (\ref{edensity}) for the term proportional to
$\Omega^{2}P^{2}$ and substitute the result into the right hand side
of the above
equation. Then, substituting this result into eqn (\ref{varN1}) gives
\begin{eqnarray}\label{varN2}
   \delta N=\integ\frac{r^{2}}{2\Omega}e^{\nu+\lambda-\phi}\left[
   \delta\rho+e^{\phi-2\lambda}\left(e^{2\lambda-2\nu}\Omega^{2}P^{2}
   +P^{\prime 2}+\left(2+\omega\right)
   e^{-\phi}\phi^{\prime 2}\right)\delta\lambda 
   +e^{\phi}P\left(e^{-2\nu}\Omega^{2}-1\right)\delta P
   \right. \nonumber \\ \left.
   -\frac{e^{\phi}}{2}\left(e^{-2\nu}\Omega^{2}P^{2}+P^{2}
   +e^{-2\lambda}P^{\prime 2}\right)\delta\phi
   -e^{\phi-2\lambda}P^{\prime}\delta P^{\prime}
   -(2+\omega)e^{-2\lambda}\phi^{\prime}
   \delta\phi^{\prime}+\delta(\Delta\phi)
   \right]\;dr.
\end{eqnarray}
Integrating the $\delta P^{\prime}$ term by parts and using the wave
equation (\ref{swave2}) one obtains an expression that cancels the $\delta P$
term  in eqn (\ref{varN2}).
Integrating the $\delta(\Delta\phi)$ term by parts we have
\begin{eqnarray}\label{deltaphi}
   \integ\frac{r^{2}}{2\Omega}e^{\nu+\lambda-\phi}\delta(\Delta\phi)\;dr
   =\integ\frac{r^{2}}{2\Omega}e^{\nu+\lambda-\phi}\left[\left(
   e^{-2\lambda}\nu^{\prime}\phi^{\prime}-e^{-2\lambda}\phi^{\prime 2}
   -\Delta\phi\right)\delta\lambda
   +e^{-2\lambda}(\phi^{\prime}-\nu^{\prime})\delta\phi^{\prime}\right]\;dr
   \nonumber \\
   +\lim_{r\to\infty}\left[\frac{e^{\nu-\phi}}{2\Omega}
   \delta\left(e^{-\lambda}r^{2}\phi^{\prime}\right)\right].
\end{eqnarray}
Substituting this result into eqn (\ref{varN2}) gives
\begin{eqnarray}\label{varN3}
   \delta N=\integ \frac{r^{2}}{2\Omega}e^{\nu+\lambda-\phi}\left[
   \delta\rho+e^{\phi}\left(e^{-2\nu}\Omega^{2}P^{2}
   +e^{-2\lambda}P^{\prime 2}+(1+\omega)e^{-2\lambda-\phi}
   \phi^{\prime 2}-e^{-\phi}\Delta\phi
   +e^{-2\lambda-\phi}\nu^{\prime}\phi^{\prime}\right)\delta\lambda 
   \right. \nonumber \\ \left.
   -\frac{e^{\phi}}{2}\left(e^{-2\nu}\Omega^{2}P^{2}+P^{2}
   +e^{-2\lambda}P^{\prime 2}\right)\delta\phi
   -e^{-2\lambda}\left(\nu^{\prime}+(1+\omega)\phi^{\prime}\right)
   \delta\phi^{\prime}\right]\;dr
   +\lim_{r\to\infty}\left[\frac{e^{\nu-\phi}}{2\Omega}
   \delta\left(e^{-\lambda}r^{2}\phi^{\prime}\right)\right].
\end{eqnarray}
Integrating the $\delta\phi^{\prime}$ term by parts and substituting
the result back into eqn (\ref{varN3}) we have
\begin{eqnarray}\label{varN4}
   \delta N=\integ \frac{r^{2}}{2\Omega}e^{\nu+\lambda-\phi}\left[
   \delta\rho+e^{\phi}\left(e^{-2\nu}\Omega^{2}P^{2}
   +e^{-2\lambda}P^{\prime 2}+(1+\omega)
   e^{-2\lambda-\phi}\phi^{\prime 2}-e^{-\phi}\Delta\phi
   \right.\right. \nonumber \\ \left.\left.
   +e^{-2\lambda-\phi}\nu^{\prime}\phi^{\prime}\right)\delta\lambda 
   -\left(\rho+\half\omega e^{-2\lambda}\phi^{\prime 2}
   -\Box\nu-\omega\Box\phi\right)
   \delta\phi\right]\;dr
   +\lim_{r\to\infty}\left[\frac{e^{\nu-\phi}}{2\Omega}
   \delta\left(e^{-\lambda}r^{2}\phi^{\prime}\right)\right]
\end{eqnarray}
where we have used eqn (\ref{edensity}) to rewrite part of the
coefficient of $\delta\phi$ in terms of the energy density $\rho$ and
rewritten some of the derivatives as the wave operator $\Box$. This
latter quantity has the explicit form
\be\label{boxdef}
   \Box f:=r^{-2}e^{-\nu-\lambda}\frac{d\;}{dr}
   \left(r^{2}e^{\nu-\lambda} f^{\prime}\right)
\ee
for any function $f(r)$. 

From the definition of the Einstein tensor we
have
\be
   G_{\alpha\beta}\xi^{\alpha}\xi^{\beta}
   =R_{\alpha\beta}\xi^{\alpha}\xi^{\beta}+\frac{{\cal R}}{2}
\ee
where $R_{\alpha\beta}$ is the Ricci tensor.
Rearranging this equation, substituting in the explicit form of
$R_{\alpha\beta}\xi^{\alpha}\xi^{\beta}=e^{-2\nu}R_{00}$
and using the definition (\ref{rho}) we have
\be
   \frac{{\cal R}}{2}=\rho-\frac{1}{r^{2}}{e^{-\nu-\lambda}}\frac{d\;}{dr}
   \left(r^{2}e^{\nu-\lambda}\nu^{\prime}\right)=\rho-\Box\nu
\ee
where we have used eqn (\ref{boxdef}) to obtain the second equality.
Solving eqn (\ref{dwave1}) for ${\cal R}$, substituting the result
into the above equation and evaluating the $\nabla^{\sigma}\phi
\nabla_{\sigma}\phi$ term gives
\be
   \rho+\half\omega e^{-2\lambda}\phi^{\prime 2}-\Box\nu-\omega\Box\phi=0.
\ee
Hence the $\delta\phi$ term in eqn (\ref{varN4}) vanishes.

Combining eqns (\ref{Gtt}), (\ref{mass1}), (\ref{rho}) and
(\ref{edensity}) one can show that
\be
   e^{2\lambda-2\nu}\Omega^{2}P^{2}
   +P^{\prime 2}-e^{2\lambda-\phi}\Delta\phi
   +e^{-\phi}\phi^{\prime}\left[\nu^{\prime}+(1+\omega)\phi^{\prime}\right]
   =\frac{2}{r}e^{-\nu-\lambda}\frac{d\;}{dr}
   \left(e^{\nu+\lambda-\phi}\right)
\ee
Substituting this result into the remaining part of eqn (\ref{varN4})
gives
\be\label{varN5}
   \delta N=\integ\frac{r^{2}}{2\Omega}\left[e^{\nu-\lambda-\phi}\delta\rho
   +\frac{\delta\lambda}{r}e^{-2\lambda}\frac{d\;}{dr}
   \left(e^{\nu+\lambda-\phi}\right)\right]\;dr
   +\lim_{r\to\infty}\left[\frac{e^{\nu-\phi}}{2\Omega}
   \delta\left(e^{-\lambda}r^{2}\phi^{\prime}\right)\right].
\ee
Taking the variation of eqns (\ref{mass1}) one can show that
\be
   \delta\rho=\frac{2}{r^{2}}\frac{d\;}{dr}\left(\delta m_{S}\right)
\ee
and
\be
   \delta\lambda=\frac{e^{2\lambda}}{r}\delta m_{S}.
\ee
Substituting both of these results into eqn (\ref{varN5}) gives
\be\label{varN6}
   \delta N=\integ\frac{1}{\Omega}\frac{d\;}{dr}\left(e^{\nu+\lambda-\phi}
   \delta m_{S}\right)\;dr   
   +\lim_{r\to\infty}\left[\frac{e^{\nu-\phi}}{2\Omega}
   \delta\left(e^{-\lambda}r^{2}\phi^{\prime}\right)\right].
\ee
Evaluating the integral and using the boundary conditions 
(\ref{bdry}) to (\ref{bdry3}) we have
\be
   \delta N=e^{-\phi_{\infty}}\lim_{r\to\infty}
   \left[\frac{\delta m_{S}}{\Omega}
   +\frac{\delta(r^{2}\phi^{\prime})}{\Omega}\right]
   =\frac{1}{\Omega}e^{-\phi_{\infty}}\delta M_{T}
\ee
where we have used eqns (\ref{Sdef1}), (\ref{mlimit1}) and (\ref{mlimit2})
to obtain the second equality. Finally, using the definition 
(\ref{MNdef}) we have
\be
   \delta M_{T}=\Omega\;\delta M_{N}.
\ee
The above equation gives the relationship between the variations in
the Tensor and Newtonian masses under the parameter change $P_{0}\rightarrow
P_{0}+\delta P_{0}$ with $\phi_{\infty}$ held fixed. Taking the limit
$\delta P_{0}\rightarrow 0$ we have
\be
   \left.\frac{dM_{T}}{dP_{0}}\right|_{\phi_{\infty}}
   =\Omega\left.\frac{dM_{N}}{dP_{0}}\right|_{\phi_{\infty}}
\ee
Hence for any $\omega$ and $\phi_{\infty}$, a solution that extremises
the Newtonian mass is also one of extremal Tensor mass. In the GR
limit, $M_{T}\rightarrow M_{ADM}$ and we recover the result derived in
\cite{jetzer}. 

\section{Conclusions}\label{concl}

We have examined three alternative definitions of mass in ST gravity:
the ADM mass $M_{ADM}$, the Keplerian mass $M_{K}$ and the Tensor mass
$M_{T}$. In a static, spherically symmetric spacetime, all three may
be written as the asymptotic limit of some quasi-local mass defined on
the hypersurfaces of constant time. In the case of the ADM mass we
have written this as the asymptotic limit of the Schwarzschild mass
$m_{S}$ which is defined in the usual way as the integral of the 
total energy density $G_{\mu\nu}\xi^{\mu}\xi^{\nu}$ over a bounded subset
of a constant time hypersurface. For a static spacetime in
GR, the Schwarzschild mass defined
in this way gives a perfectly satisfactory notion of quasi-local
mass. We have analysed the behaviour of
the asymptotic and quasi-local masses for BD boson stars in the strong
coupling ($\omega=-1$) case and found that both $m_{S}$ and $M_{ADM}$ have
undesirable physical properties even though the boson field energy
momentum tensor obeys the weak energy condition. In addition, for
strong couplings, $M_{ADM}<M_{N}$ for all values of the central boson
field amplitude. If one assumes that the energy of the star is given
by $M_{ADM}$, one would have to conclude that all stars in the strong
coupling case are energetically stable, regardless of their central
density. This contrasts strongly with the behaviour of boson star
solutions in both GR and in weakly coupled BD theory. In all
of these cases, one finds that the solutions become unstable above a 
certain value of the central density. These results imply that
$M_{ADM}$ does not give a correct description of the energy of the
spacetime, primarily because it does not correctly include the
dilaton field's contribution to the total energy. The dilaton, being
an extra component of the gravitational field, must be treated on a
slightly different footing to the normal matter fields.

Both $M_{T}$ and $M_{K}$ adequately include the contribution made by
the dilaton to the active gravitational mass of the boson star as seen
by two different kinds of orbiting test particles. However, as pointed
out by Lee \cite{lee1}, out of the range of possible masses one can 
derive from the superpotential (\ref{spotential}), it is only the
Tensor mass that is conserved in a non-static isolated source. 
This implies that the Tensor mass may be the true physical energy 
of the star. However, the analysis given by Lee does not show 
that $M_{T}$ uniquely satisfies a conservation law: one can
conceive of other definitions of energy that are also conserved. In
this work we have shown that, for a one parameter set of boson star 
solutions in BD theory, $M_{T}$ is unique in that it is the
only conserved quantity whose extremal points coincide with those of
the boson particle number. In this respect, $M_{T}$ is the
analogue of the ADM energy used in GR and this behaviour 
supports the conjecture that $M_{T}$ uniquely describes the 
energy of the system. We see no reason why these results should not
hold for stellar objects composed of matter other than bosons, or in
more general scalar-tensor theories.

For the weak coupling case, the numerical difference between $M_{T}$
and $M_{ADM}$ is small, so in practice there is little accuracy lost in
identifying the energy of a boson star with $M_{ADM}$. However, in the
early Universe and in more general scalar-tensor theories, one may
have to consider strong couplings and in these cases it is important 
that the correct definition of mass is used.

\subsection*{Acknowledgements}

I would like to thank Andrew Liddle, James Lidsey, Malcolm MacCallum, 
Franz Schunck and Diego Torres for helpful conversations relating to this work.

\newpage

\begin{figure}
\centering 
\leavevmode\epsfxsize=8cm \epsfbox{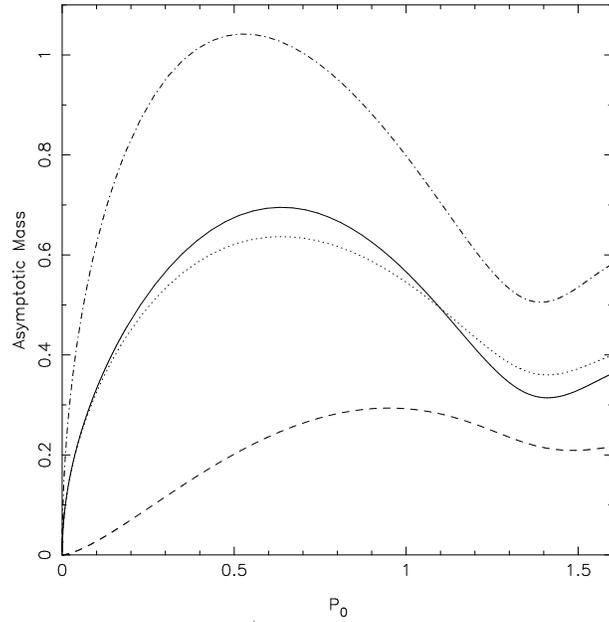}\\ 
\caption{Mass curves for the $\omega=-1$ solutions with 
$e^{\phi_{\infty}}=1$. The curves are parameterised 
by $P_{0}$, the amplitude of the boson wave
function at the origin. Four mass curves are shown: the Newtonian mass 
$M_{N}$ (solid line), the ADM mass $M_{ADM}$ (dashed line),
the Tensor mass $M_{T}$ (dotted line) and the 
Keplerian mass $M_{K}$ (dot-dashed line). All masses are 
expressed in units of $M_{pl}^{2}/\alpha$.}
\end{figure} 

\end{document}